\documentclass[10pt, preprint2]{emulateapj}
\usepackage{verbatim}
\usepackage[usenames, dvipsnames]{color}
\usepackage[pagebackref]{hyperref}

\newcommand{\target}{$\pi$~Men}
\newcommand{\thisstar}{$\pi$~Men}
\newcommand{\thisplanetb}{$\pi$~Men~b$ $}
\newcommand{\thisplanetc}{$\pi$~Men~c$ $}

\newcommand{\rsun}{\ensuremath{R_\sun}}
\newcommand{\msun}{\ensuremath{M_\sun}}
\newcommand{\lsun}{\ensuremath{L_\sun}}

\newcommand{\rearth}{\ensuremath{R_\earth}}
\newcommand{\mearth}{\ensuremath{M_\earth}}

\newcommand{\mjup}{\ensuremath{M_{\rm J}}}

\newcommand{\rpl}{\ensuremath{R_{p}}}
\newcommand{\mpl}{\ensuremath{M_{p}}}

\newcommand{\rhopl}{\ensuremath{\rho_{p}}}

\newcommand{\rstar}{\ensuremath{R_\star}}
\newcommand{\mstar}{\ensuremath{M_\star}}
\newcommand{\lstar}{\ensuremath{L_\star}}

\newcommand{\teffstar}{\ensuremath{T_{\rm eff}}}
\newcommand{\rhostar}{\ensuremath{\rho_\star}}
\newcommand{\loggstar}{\ensuremath{\log{g}}}

\newcommand{\arstar}{\ensuremath{a/\rstar}}

\newcommand{\kms}{km~s$^{-1}$}

\newcommand{\Kepler}{\emph{Kepler}}
\newcommand{\TESS}{\emph{TESS}}
\newcommand{\HARPS}{HARPS}

\newcommand{\gaia}{\emph{Gaia}}

\newcommand{\ms}{\ensuremath{\rm m\,s^{-1}}}

\newcommand{\gcmc}{\ensuremath{\rm g\,cm^{-3}}}
\newcommand{\ergscmsq}{\ensuremath{\rm erg\,s^{-1}\,cm^{-2}}}

\newcommand{\tessfitP}{\ensuremath{6.2679\pm0.00046}}
\newcommand{\tessfitTc}{\ensuremath{2458325.50400_{-0.00074}^{+0.0012}}}
\newcommand{\tessfitRratio}{\ensuremath{0.01703_{-0.00023}^{+0.00025}}}
\newcommand{\tessfitb}{\ensuremath{0.59_{-0.020}^{+0.018}}}
\newcommand{\tessfitaor}{\ensuremath{13.38\pm0.26}}
\newcommand{\tessfitinc}{\ensuremath{87.456_{-0.076}^{+0.085}}}
\newcommand{\tessfitRp}{\ensuremath{2.042\pm0.050}}
\newcommand{\tessfitTdur}{\ensuremath{2.953\pm0.047}}
\newcommand{\tessfitsemi}{\ensuremath{0.06839\pm0.00050}}
\newcommand{\tessfitTeq}{\ensuremath{1169.8_{-4.3}^{+2.8}}}
\newcommand{\tessfitFinso}{\ensuremath{0.42_{-0.09}^{+0.04}}}

\newcommand{\tessfitMp}{\ensuremath{4.82_{-0.86}^{+0.84}}}
\newcommand{\tessfitrhop}{\ensuremath{2.97_{-0.55}^{+0.57}}}
\newcommand{\tessfitloggp}{\ensuremath{3.041_{-0.86}^{+0.07}}}

\newcommand{\AATfittkb}{\ensuremath{192.6\pm1.4}}
\newcommand{\AATfittsecoswb}{\ensuremath{0.6957\pm0.0044}}
\newcommand{\AATfitPb}{\ensuremath{2093.07\pm1.73}}
\newcommand{\AATfittpb}{\ensuremath{2445852.0\pm3.0}}
\newcommand{\AATfittcb}{\ensuremath{2446087.0\pm8.4}}
\newcommand{\AATfittsesinwb}{\ensuremath{-0.392\pm0.006}}
\newcommand{\AATfiteb}{\ensuremath{0.637\pm0.002}}
\newcommand{\AATfitwb}{\ensuremath{330.61\pm0.3}}
\newcommand{\AATfittkc}{\ensuremath{1.58_{-0.28}^{+0.26}}}

\shorttitle{Transiting Super-Earth in $\pi$ Mensae}
\shortauthors{}

\begin{document}

\title{TESS Discovery of a Transiting Super-Earth in the $\pi$ Mensae System}

\author{Chelsea X.\ Huang\altaffilmark{1,2}, 
Jennifer Burt\altaffilmark{1,2},
Andrew Vanderburg\altaffilmark{3,4},
Maximilian N.\ G{\"u}nther\altaffilmark{1,2},
Avi Shporer\altaffilmark{1},
Jason A.\ Dittmann\altaffilmark{5,6},
Joshua N.\ Winn\altaffilmark{7}, 
Rob Wittenmyer\altaffilmark{8},
Lizhou Sha\altaffilmark{1},
Stephen R.\ Kane\altaffilmark{9},
George R.\ Ricker\altaffilmark{1},
Roland K.\ Vanderspek\altaffilmark{1},
David W.\ Latham\altaffilmark{10},
Sara Seager\altaffilmark{1,6},
Jon M.\ Jenkins\altaffilmark{11},
Douglas A.\ Caldwell\altaffilmark{11,12}
Karen A.\ Collins\altaffilmark{10},
Natalia Guerrero\altaffilmark{1},
Jeffrey C.\ Smith\altaffilmark{12},
Samuel N.\ Quinn\altaffilmark{11},
St\'ephane Udry\altaffilmark{13}, 
Francesco Pepe\altaffilmark{13}, 
Fran\c{c}ois Bouchy\altaffilmark{13}, 
Damien S\'egransan\altaffilmark{13}, 
Christophe Lovis\altaffilmark{13}, 
David Ehrenreich\altaffilmark{13}, 
Maxime Marmier\altaffilmark{13}, 
Michel Mayor\altaffilmark{13},
Bill Wohler\altaffilmark{11,12},
Kari Haworth\altaffilmark{1},
Edward H.\ Morgan\altaffilmark{1},
Michael Fausnaugh\altaffilmark{1},
David R. Ciardi\altaffilmark{14},
Jessie Christiansen\altaffilmark{14},
David Charbonneau\altaffilmark{10},
Diana Dragomir\altaffilmark{1,15},
Drake Deming \altaffilmark{16},
Ana Glidden\altaffilmark{1,6},
Alan M.\ Levine\altaffilmark{1},
P.R.\ McCullough \altaffilmark{17},
Liang Yu\altaffilmark{1},
Norio Narita\altaffilmark{18,19,20,21,22},
Tam Nguyen\altaffilmark{1},
Tim Morton\altaffilmark{7},
Joshua Pepper\altaffilmark{23},
Andr\'as P\'al \altaffilmark{1,24,25},
Joseph E.\ Rodriguez \altaffilmark{10},
and the TESS team}

\altaffiltext{1}{Department of Physics, and Kavli Institute for Astrophysics and Space Research, Massachusetts Institute of Technology, Cambridge, MA 02139, USA}
\altaffiltext{2}{Juan Carlos Torres Fellow}
\altaffiltext{3}{Department of Astronomy, The University of Texas at Austin, Austin, TX 78712, USA}
\altaffiltext{4}{NASA Sagan Fellow}
\altaffiltext{5}{51 Pegasi b Postdoctoral Fellow}
\altaffiltext{6}{Earth and Planetary Sciences, MIT, 77 Massachusetts Avenue, Cambridge, MA 02139, USA}
\altaffiltext{7}{Department of Astrophysical Sciences, Princeton University, 4 Ivy Lane, Princeton, NJ 08544, USA}
\altaffiltext{8}{University of Southern Queensland, West St, Darling Heights QLD 4350, Australia}
\altaffiltext{9}{Department of Earth Sciences, University of California, Riverside, 
CA 92521, USA}
\altaffiltext{10}{Harvard--Smithsonian Center for Astrophysics, Harvard University, Cambridge, MA 02138, USA}
\altaffiltext{11}{NASA Ames Research Center, Moffett Field, CA, 94035} 
\altaffiltext{12}{SETI Institute, Mountain View, CA 94043, USA}
\altaffiltext{13}{Observatoire de l'Universit\'e de Gen\`eve, 51 chemin des Mail-lettes, 1290 Versoix, Switzerland}
\altaffiltext{14}{NASA Exoplanet Science Institute,	Caltech/IPAC-NExScI	1200 East California Boulevard Pasadena, CA 91125}
\altaffiltext{15}{NASA Hubble Fellow}
\altaffiltext{16}{Department of Astronomy, University of Maryland at College Park, College Park MD, 20742}
\altaffiltext{17}{Department of Physics and Astronomy, Johns Hopkins University, 3400 North Charles Street, Baltimore, MD 21218, USA}
\altaffiltext{18}{Department of Astronomy, The University of Tokyo, 7-3-1 Hongo, Bunkyo-ku, Tokyo
113-0033, Japan}
\altaffiltext{19}{Astrobiology Center, National Astronomical Observatory of Japan, NINS, 2-21-1 Osawa, Mitaka, Tokyo 181-8588, Japan}
\altaffiltext{20}{JST, PRESTO, 7-3-1 Hongo, Bunkyo-ku, Tokyo 113-0033, Japan}
\altaffiltext{21}{Instituto de Astrof\'{i}sica de Canarias (IAC), 38205 La Laguna, Tenerife, Spain}
\altaffiltext{22}{National Astronomical Observatory of Japan, NINS, 2-21-1 Osawa, Mitaka, Tokyo 181-8588, Japan}
\altaffiltext{23}{Department of Physics, Lehigh University, 16 Memorial Drive East, Bethlehem, PA 18015, USA}
\altaffiltext{24}{Department of Astronomy, Lor\'and E\"{o}tv\"{o}s University,
P\'azm\'any P. stny. 1/A, Budapest H-1117, Hungary}
\altaffiltext{25}{Konkoly Observatory, Research Centre for Astronomy and
Earth Sciences, Hungarian Academy of Sciences, Konkoly Thege
Mikl\'os \'ut 15-17, H-1121 Budapest, Hungary}
\begin{abstract}

We report the detection of a transiting planet around
\target\ (HD\,39091), using data from the \textit{Transiting Exoplanet Survey Satellite} (\TESS).  The solar-type host star is unusually bright ($V=5.7$) and was already known to host a Jovian planet on a highly eccentric, 5.7-year orbit.  The newly discovered planet has a
size of $2.04\pm 0.05$~$R_\oplus$ and an orbital period of 6.27 days.
Radial-velocity data from the HARPS and AAT/UCLES archives also displays a 6.27-day periodicity, confirming the existence of the planet and leading to a mass determination of $4.82\pm
0.85$~$M_\oplus$.  The star's proximity and brightness will facilitate further investigations, such as atmospheric spectroscopy, asteroseismology, the Rossiter--McLaughlin effect, astrometry, and direct imaging.
\end{abstract}

\keywords{planetary systems, planets and satellites: detection, stars: individual (HD 39091, TIC 261136679)}
\section{Introduction}
\label{sec:intro}

The mission of the {\it Transiting Exoplanet Survey Satellite} \citep[\TESS,][]{ricker} is to search for transiting planets as small as Earth around the nearest and brightest stars.  Four 10\,cm optical telescopes are used to repeatedly image wide fields and monitor the brightness of suitable stars. The data are then searched for periodic dips that could be caused by transiting planets. The spacecraft was launched on April 18, 2018 and began the sky survey on July 25.  Here, we report on the discovery of a small transiting planet around a bright star \thisstar. 

\thisstar\ (also known as HD~39091) is a naked-eye G0V star at a distance of 18.27{\bf $\pm$0.02}~pc \citep{GaiaDR2} with a mass of $1.1 \, M_\odot$ and a radius of $1.1 \, R_\odot$.  Doppler monitoring by \cite{Jones2002} and \cite{Wittenmyer2012}
revealed a planet (\thisplanetb) with a mass about 10 times that of Jupiter,
an orbital period of 5.7~years, and an orbital eccentricity of 0.6.
With a visual apparent magnitude of 5.67, the star is a prime target
for the \TESS\ survey. It is one of several hundred thousand
pre-selected stars for which data will be available with 2-minute time sampling,
as opposed to the 30-minute sampling of the full image data set.

This \emph{Letter} is organized as follows.
Section~\ref{sec:data} presents
the \TESS{} photometric data that led to the detection of the new
planet \thisplanetc, as well as the archival radial velocity
data that confirm the planet's existence.
Section~\ref{sec:analysis} describes our methods for determining the system parameters,
including the mass and radius of the star and planet.
Section~\ref{sec:discussion} discusses 
some possible follow-up observations that will be facilitated
by the star's brightness and proximity to Earth.


\section{Observations and Data Reduction}
\label{sec:data}

\subsection{\TESS\ photometry}

The \TESS\ survey divides the sky into 26 partially overlapping sectors,
each of which is observed for approximately one month during the two-year primary mission.
\thisstar\ is located near the southern ecliptic pole in a region where 6 sectors overlap,
implying that it is scheduled to be observed for a total of 6 months.
This paper is based on data from Sector 1 (2018 July 25 -- August 22), during
which \thisstar\ was observed with CCD~2 of Camera~4.

The data were processed with two independently written codes:
the MIT Quick Look Pipeline (partially based on {\tt fitsh}, \citet{Pal:2009}),
which analyzes the full images that are obtained with
30-minute time sampling; and the
Science Processing Operations Center pipeline, a descendant
of the \Kepler\ mission pipeline based at the NASA Ames Research Center \citep{jenkins2010},
which analyzes the 2-minute data that are obtained for pre-selected target stars.
For \thisstar, both pipelines detected a signal with a
period of 6.27 days, an amplitude of about 300~ppm, a duration of 3 hours,
and a flat-bottomed shape consistent with
the light curve of a planetary transit.

Previous surveys taught us that transit-like signals sometimes turn out to
be eclipsing binaries that are either grazing, or blended with a bright star,
causing the amplitude of the signal to be deceptively small and resemble that
of a planet \citep[e.g.][]{Cameron2012}.
In this case, the signal survived all the usual tests for such ``false positives.''
There is no discernible secondary eclipse, no detectable alternation
in the depth of the transits, and no detectable motion
of the stellar image on the detector during the fading
events.\footnote{The last test in the
list, the centroid test, was complicated by the fact that the star is
bright enough to cause blooming in the \TESS\ CCD images. The associated
systematic effects were removed using the method of \cite{Guenther2017b}.}

\begin{figure}
\centering
	\includegraphics[width=\columnwidth]{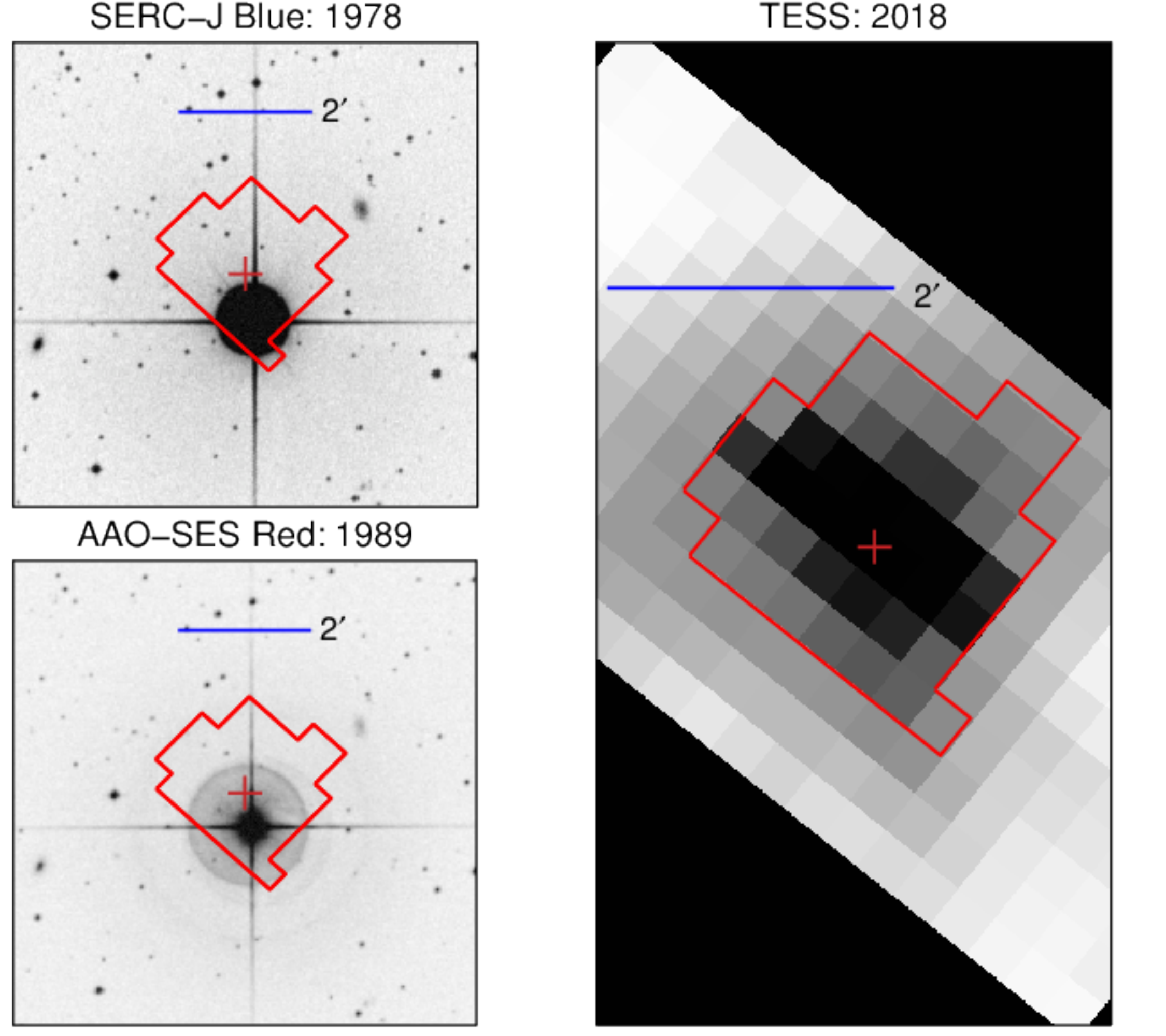}
    \caption{Images of the field surrounding \thisstar.
    \textit{Top left.}---From the
    Science and Engineering Research Council J survey, obtained with
    a blue-sensitive photographic emulsion in 1978. The red cross is the current position of \thisstar. Red lines mark the boundary
    of the \TESS{} photometric aperture.
    \textit{Bottom left.}---From the AAO
    Second Epoch Survey, obtained with a red-sensitive photographic emulsion
    in 1989.
    \textit{Right.}---Summed \TESS{} image. North is up and East is to the left in all the images.} 
    \label{fig:imaging}
\end{figure}

After identifying the transits, we tried improving
on the light curve by experimenting with different
choices for the photometric aperture, including circles as well
as irregular pixel boundaries
that enclose the blooming stellar image.
Best results were obtained for
the aperture shown in Figure~\ref{fig:imaging}. Also shown are
images of the field from optical sky surveys conducted 30--40 years ago,
long enough for the star to have moved about an arcminute relative to the
background stars. This allows us a clear view along the line of sight
to the current position of \thisstar, which is reassuringly blank:
another indication that the transit signal is genuine and not an unresolved
eclipsing binary. The other stars within the photometric
aperture are too faint to cause the 300~ppm fading events.

The top panel of Figure~\ref{fig:transitplot} shows the result of simple
aperture photometry.  Most of the observed variation is instrumental.
There may also be a contribution from stellar variability, which is
expected to occur on the 18-day timescale of the rotation period \citep{Zurlo+2018}.
To remove these variations and permit a sensitive search for transits,
we fitted a basis spline with knots spaced by 0.3 days, after
excluding both 3$\sigma$ outliers and the data obtained during and immediately
surrounding transits. We then divided the light curve by the best-fitting spline.

The middle panel of Figure~\ref{fig:transitplot} shows the result.
The scatter is 142~ppm per 2-minute sample, and 30~ppm
when averaged into 6-hour bins, comparable
to the highest-quality \Kepler\ light curves.
The gap in the middle of the time series occurred when
observations were halted for data
downlink.
The other gap occurred during
a period when the spacecraft pointing
jitter was higher than normal.
We also excluded the data from the 30--60 minute intervals surrounding
``momentum dumps,'' when thrusters are fired to reorient
the spacecraft and allow the reaction wheels to spin down. 
The times of the momentum dumps are marked in Figure~\ref{fig:transitplot}.
There were
10 such events during Sector 1 observations, occurring every 2 and half days.

\subsection{Radial-velocity data}

\target\ has been monitored for 20 years as part of the Anglo-Australian Planet Search, which uses the 3.9m Anglo-Australian Telescope (AAT) and the University College London Echelle Spectrograph \citep[UCLES;][]{Diego:1990}.
The long-period giant planet \thisplanetb\ was discovered in this survey \citep{Jones2002, Butler2006}. A total of 77 radial velocities are available,
obtained between 1998 and 2015, with a mean internal uncertainty of 2.13~\ms. 

The star was also monitored with the
High-Accuracy Radial-velocity Planet Searcher \HARPS\
\citep{Mayor2003} on the ESO~3.6m telescope at
La Silla Observatory in Chile.  A hardware upgrade in June 2015
led to an offset in the velocity scale \citep{LoCurto2015}.
For this reason, our model allows for different constants to be added to the pre-upgrade and post-upgrade data.
A total of 145 radial velocities are available, obtained between December 2003 and March 2016 with irregular sampling.
The mean internal uncertainty of the 128 pre-upgrade velocities is 0.78~\ms,
while that of the 17 post-upgrade
velocities is 0.38~\ms.

The top panel of Figure~\ref{fig:rvplot} shows the radial-velocity data.
It is easy to see the 400~\ms\ variations from the giant planet.
To search for evidence of the new planet, we subtracted the best-fitting
single-planet model from the data and computed the Lomb--Scargle
periodogram of the more precise HARPS data, shown in the middle panel.  The highest peak is far above
the 0.1\% false alarm threshold and is located at the transit
period of 6.27~days. The next highest peaks, bracketing a period of 1 day, are
aliases of this signal.  The phase of the 6.27-day signal is also consistent with the measured transit times.

We consider this to be a decisive confirmation of the existence of \thisplanetc.
Still, as another precaution against false positives,
we checked the \HARPS\ spectra for any indication
of a second star, or spectral-line distortions associated with the 6.27-day signal.
We re-analyzed the HARPS cross-correlation functions with the {\scshape blendfitter} routine \citep{Guenther2018} and found no sign of any correlated bisector variations.

\section{Determination of System Parameters}
\label{sec:analysis}

\begin{figure*}
\includegraphics[width=\linewidth]{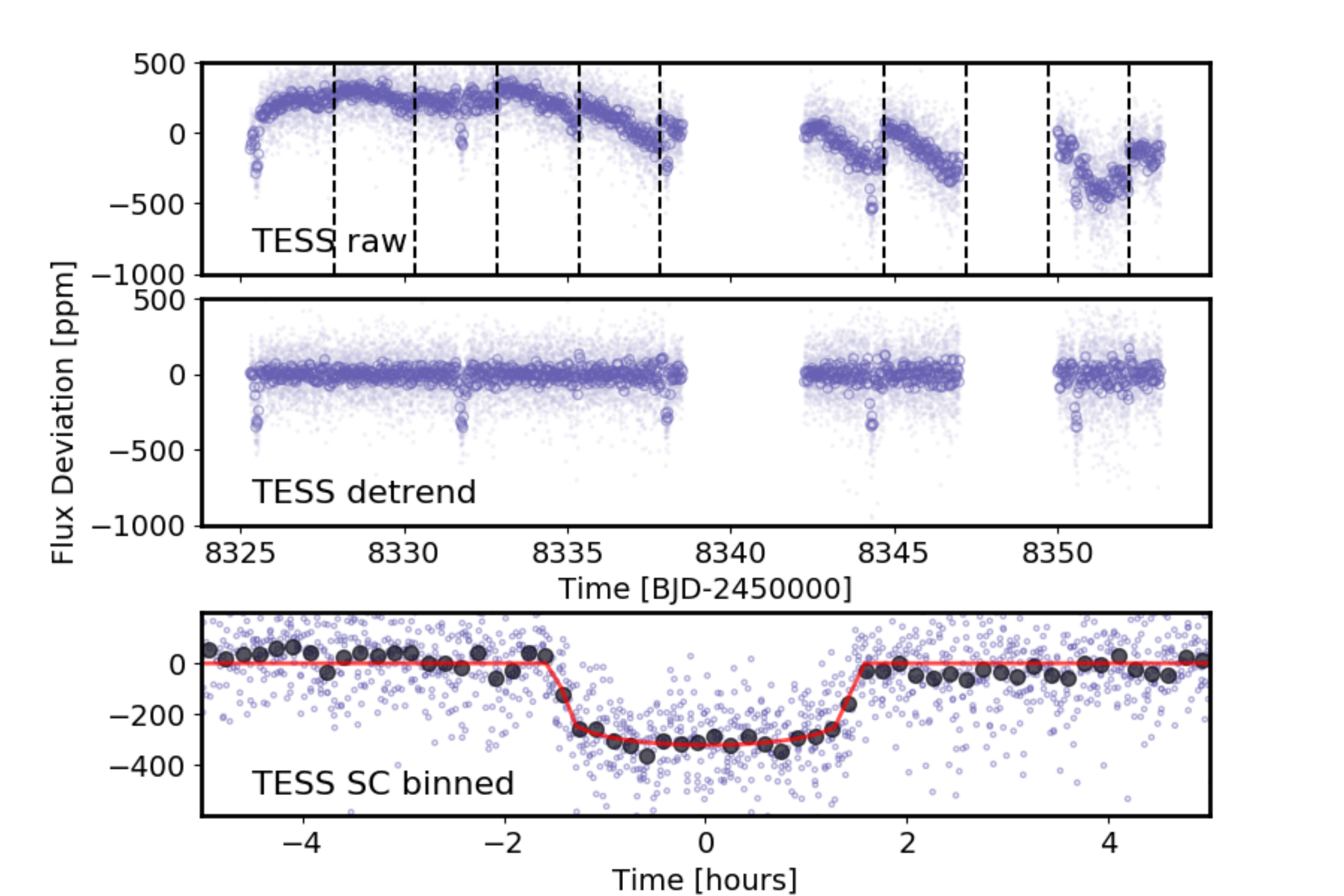}
\caption{Raw (\emph{top}) and corrected (\emph{middle}) \TESS{} light curves.
The lighter points are based on the short cadence (SC)
data with 2-minute sampling.
The darker points are 30-minute averages.
The dashed lines indicate the times of momentum dumps.
The interruptions are from the data
downlink and the pointing anomaly. The bottom panel
shows the phase-folded light curve, along with
the best-fitting model.
The black dots represent 5-minute averages.
\label{fig:transitplot}}
\vspace{0.7cm}
\end{figure*}

\begin{figure*}
\centering
	\includegraphics[width=\linewidth]{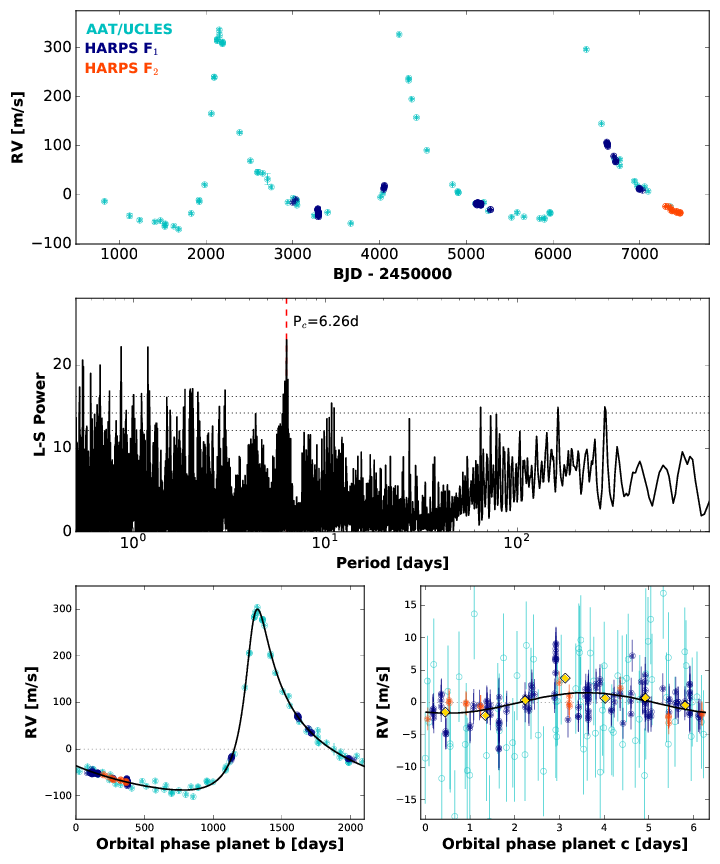}
    \caption{\textit{Top.}---Relative radial
    velocity of \thisstar\ as measured
    with UCLES and HARPS (both pre- and post-upgrade).
    The zero points of each of the 3 datasets have
    been adjusted to coincide.
    \textit{Middle.}---Lomb--Scargle periodogram of the \HARPS\ data,
    after subtracting the
    single-planet model
    that best fits the entire data set.
    The dotted lines are the power levels
    corresponding to false alarm probabilities of
    10\%, 1\%, and 0.1\%.
    \textit{Bottom left.}---Radial velocity as a function
    of the orbital phase of planet b, after subtracting
    the best-fitting model for the variation due to planet c.
    \textit{Bottom right.}---Similar, but for planet c. The orange point is binned in phase space. 
    \label{fig:rvplot}}
\end{figure*}

\begin{figure*}
\includegraphics[width=\linewidth]{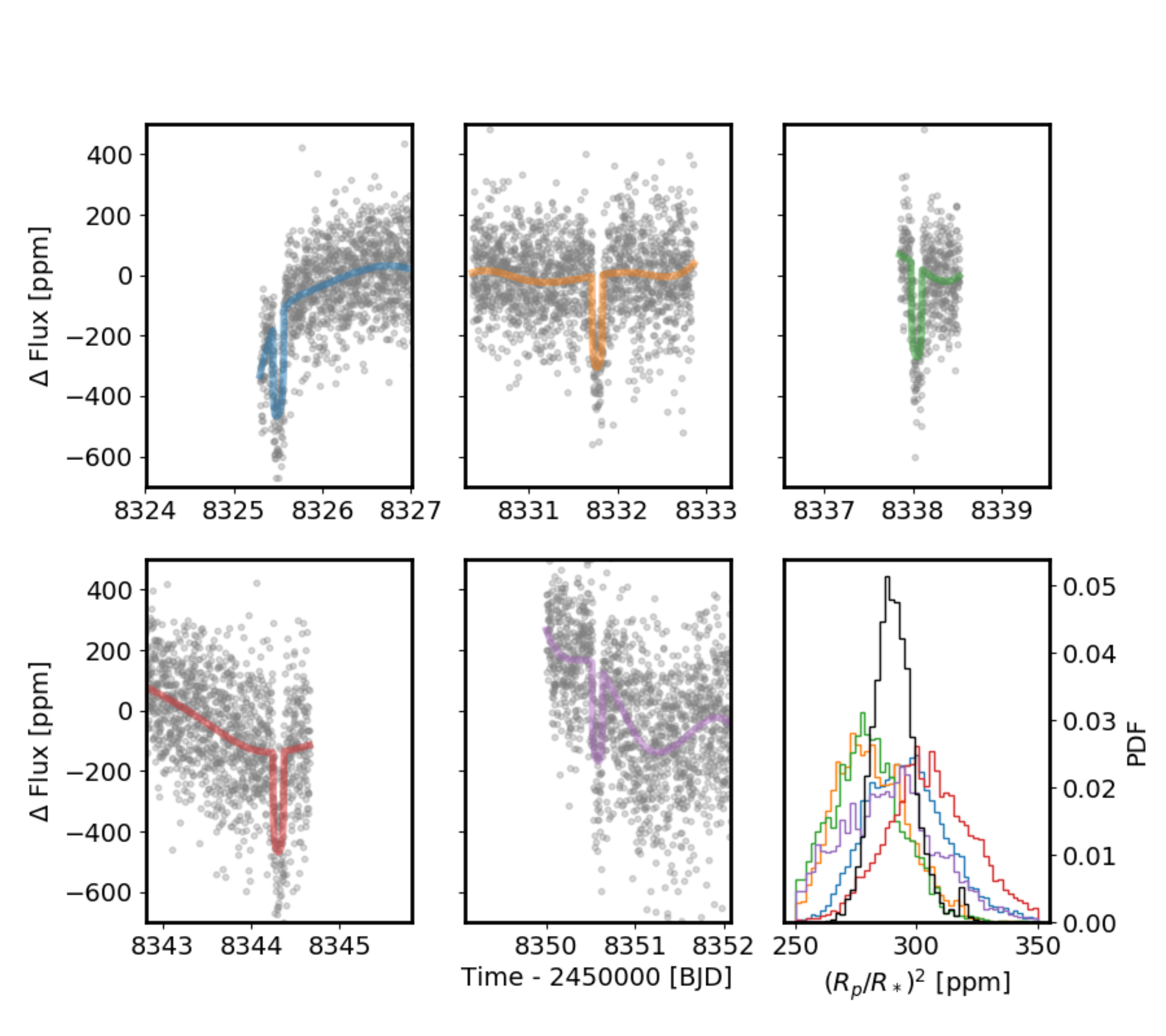}
\caption{Plotted are the \TESS{} data that surround each of the five
observed transits and were obtained in between momentum dumps.
Each panel shows 3 days of data and spans the same range
of flux deviations.
In the bottom right panel, the colored histograms are
the 5 posterior distributions for $(R_p/R_*)^2$, obtained from independent fits to the 5 transit datasets. The black histogram
is the posterior based on the fit to all the data.
\label{fig:mcmc}}
\end{figure*}

We performed a joint analysis of the two-planet system using the \TESS{} transit light curve and the 222 radial velocities from the AAT and HARPS surveys. 
The orbit of planet c was assumed
to be circular in the fit.\footnote{We also tried allowing planet c to have an eccentric orbit, which resulted in an upper limit of $e_c < 0.3$ (1$\sigma$).
{\bf All of the other orbital parameters remained consistent 
with the results of the $e_c\equiv 0$ model,
although naturally, some parameters were subject
to slightly larger uncertainties.}
}
As noted previously, we assigned a different additive
constant to each of the 3 radial-velocity data sets.
We also allowed for
3 independent values of the ``jitter'', a term that is added in quadrature to the internally-estimated
measurement uncertainty to account for systematic effects.

We assumed the star to follow a quadratic limb-darkening
law and used the formulas of \citet{MandelAgol:2002} as
implemented by \citet{Kreidberg(2015)}.
We fixed the limb-darkening coefficients
at $u_1 = 0.28$ and $u_2= 0.27$, based on the tabulation of \citet{Claret:2018}.
The photometric model was computed with 0.4~min sampling
and then averaged to 2~min
before comparing with the data.

We also fitted for the mass and radius of the star,
which were constrained by measurements of the spectroscopic
parameters \citep{Ghezzi2010} as well
as the stellar mean density $\rho_\star$ implicit in the
combination of $P$, $a/R_\star$, and $i$ \citep{SeagerMallenOrnelas2003,Winn2010}.
For a given choice of mass, age, and metallicity,
we relied on the Dartmouth stellar-evolutionary
models \citep{Dotter:2008} to determine the corresponding
radius $R_\star$, effective temperature $T_{\rm eff}$,
and \gaia{} absolute magnitude.
The likelihood function enforced agreement
with the measurements of $T_{\rm eff}$, $\rho_\star$, $\log g$,
and parallax (based on the absolute and apparent \gaia{} magnitudes).

To determine the credible intervals for all the parameters,
we used the Markov Chain Monte Carlo (MCMC) method as implemented
in {\tt emcee} by \citet{ForemanMackey:2012}. 
Detrending was performed simultaneously with the transit
    fitting: at each step in the Markov Chain, the transit model ({\tt batman}, \citet{Kreidberg(2015)})
was subtracted from the data and the residual light
curve was detrended using a basis spline with knots spaced by 0.5 days,
To avoid trying to model the discontinuities in the data
related to momentum dumps, we only fitted the segment
of the light curve in between momentum dumps.
The results are given in Table~\ref{tab:stellar} and Table~\ref{tab:planet},
and the
best-fitting model is plotted in Figures~\ref{fig:transitplot} and
\ref{fig:rvplot}.
As a consistency check, we also fitted each of the 5
transits independently. Figure~\ref{fig:mcmc} shows
the results, which are all consistent to within
the estimated uncertainties.

\begin{figure*}
\centering
	\includegraphics[width=0.95\linewidth]{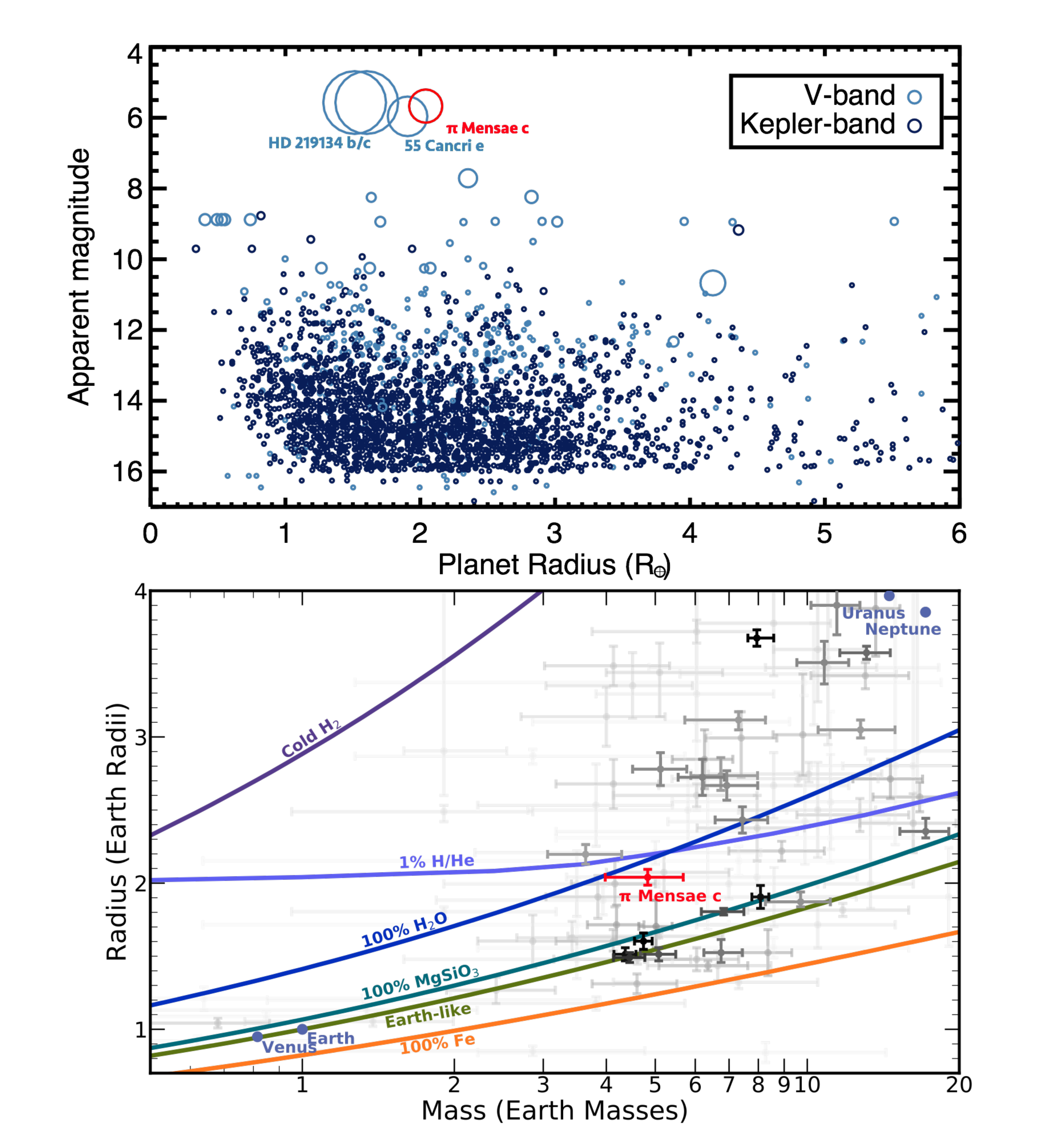}
    \caption{\thisplanetc\ in the context of other known exoplanets. \textit{Top.}---Apparent magnitude and planet radius for all
    the known transiting planets.
    The $V$ magnitude is plotted when available, and otherwise the \Kepler{}
    magnitude is plotted. The symbol size is proportional to the angular diameter of the star.
    \textit{Bottom.}---Mass-radius diagram for small exoplanets. Darker points represent more precise measurements.  Based on data from the NASA Exoplanet Archive, accessed on 13 September 2018.\footnote{\url{https://exoplanetarchive.ipac.caltech.edu/cgi-bin/TblView/nph-tblView?app=ExoTbls\&config=planets}} 
    Model curves are: $H_2$ \citep{Seager:2007}; 100\% $H_2O$, 100\% $M_gS_iO_3$, 100\% $F_e$, Earth like \citep[]{Zeng:2016}; and 1 \% H/$H_e$\citep{Lopez:2012}. \label{fig:dis}} 
\end{figure*}

\section{Discussion}
\label{sec:discussion}

Among the known stars with transiting planets, \thisstar\ is the second brightest in the visual band, as illustrated in the top panel
of Figure~\ref{fig:dis}.  \TESS\ has begun to fulfill its promise
to enlarge the collection of small, transiting planets orbiting bright stars.
Such stars enable precise measurements of
that planet's mass and radius.
The bottom panel of Figure~\ref{fig:dis}
shows the measured masses and radii of the known planets
smaller than Neptune, overlaid with
theoretical mass/radius relationships for different compositions.
\thisplanetc\ falls above the ``pure rock" curve on the diagram, and near curves for planets composed of either pure water or rocky interiors surrounded by a lightweight 1\% H/He envelope. \thisplanetc\ must not have a purely rocky composition, but instead may have a rocky core surrounded by layers of volatiles, such as hydrogen/helium (see \citet{OwenWu:2017}), or water/methane \citep{Vanderburg:2017}. 


With a near-infrared magnitude of $K = 4.24$, \thisstar\ is also
one of the brightest stars available for
planetary atmospheric characterization
with the \emph{James Webb Space Telescope} (\emph{JWST}). 
\thisplanetc\ is one of the top 10 most favorable systems in the ranking
scheme of \citet{Kempton:2018}, although this ranking scheme
does not take into account the practical difficulties in achieving
photon-limited observations of such a bright star.
Transit spectroscopy would be difficult if the planet has
an Earth-like atmospheric scale height of order 10~km, in which case
the atmospheric signals would be on the order of only 1~ppm.
On the other hand, given the intense stellar irradiation, there may be larger signals
form an escaping atmosphere\citep[see, e.g.,][]{Ehrenreich:2015,Spake:2018}.
Spectroscopy of occultations (secondary eclipses) is also promising.
The occultation depth is predicted to be 60~ppm in the
Rayleigh-Jeans limit, assuming the entire surface radiates as a blackbody at the equilibrium temperature of 1200~K.

Another interesting possibility is
to measure the stellar obliquity by observing the Rossiter--McLaughlin (RM) effect.
Stars with close-in giant planets show
a surprising diversity of orientations \citep{WinnFabrycky2015,Triaud2017}.
However, we know relatively little
about the obliquities of stars with
smaller planets, because the
relevant signals are smaller and harder to detect.
In the case of \thisplanetc,
the amplitude of the RM effect is on the order of 1~\ms, the product of the transit depth (300~ppm) and the sky-projected
rotation velocity (3.1~km~s$^{-1}$; \citealt{ValentiFischer2005}).

The \thisstar\ system consists of a giant
planet on a long-period, highly eccentric
orbit, along with a planet with an orbit and mass
that are both smaller by two orders of magnitude.
Recent follow-up studies of {\it Kepler} systems
have suggested that they maybe intrinsically common \citep{Bryan2018,Zhu2018}.
Thus, we might find many similar cases
with \TESS, providing clues about the formation of close-orbiting planets,
whether by disk migration, Lidov--Kozai oscillations, or other mechanisms.

Astrometric observations with the \gaia{} spacecraft might
ultimately reveal the full three-dimensional geometry of the system.
\citet{Ranalli:2017} predicted that the
astrometric signal of \thisplanetb\ will be detectable
with a signal-to-noise ratio higher than 10 by the end of the mission.
Indeed, the fit to the existing \gaia{} data
exhibits an excess scatter of $295 \, \mu\arcsec$ ($37\sigma$),
perhaps a hint of planet-induced motion.
Direct imaging might also be fruitful some day, although
\cite{Zurlo+2018} have already ruled out any 
companions with orbital separation 10--20 AU and
an infrared contrast exceeding $10^{-6}$, corresponding
roughly to 30 Jupiter masses.

While some of these observations may be far off, we will not have
to wait long for another
opportunity to learn more about \thisstar.
As mentioned earlier, \TESS\ is scheduled to collect 5 additional months of data.
This will allow us to refine our knowledge of planet c, search for additional transiting planets, and try to detect 
asteroseismic oscillations.
The \thisstar\ system has already been generous
to the exoplanet community, and with a little luck, the gifts
will keep arriving.

\begin{deluxetable*}{lcr}
\tablewidth{0pc}
\tabletypesize{\scriptsize}
\tablecaption{
    System Parameters for \target
    \label{tab:stellar}
}

%
\startdata
\hline\hline
\multicolumn{1}{l}{\bf Stellar Parameters} & \multicolumn{1}{c}{\bf Value} & \multicolumn{1}{r}{{\bf Source}} \\
\noalign{\vskip -3pt}
\sidehead{Catalog Information}
~~~~R.A. (h:m:s)                      &   05:37:09.89 & \gaia{} DR2\\
~~~~Dec. (d:m:s)                      &  $-80$:28:08.8   & \gaia{} DR2\\
~~~~Epoch							  &  2015.5         & \gaia{} DR2 \\
~~~~Parallax (mas)                    & $54.705\pm0.067$ & \gaia{} DR2\\
~~~~$\mu_{\mathrm{ra}}$ (mas yr$^{-1}$)        & $311.19\pm0.13$  & \gaia{} DR2 \\
~~~~$\mu_{\mathrm{dec}}$ (mas yr$^{-1}$)       & $1048.85\pm0.14$ & \gaia{} DR2\\
~~~~\gaia{} DR2 ID                       &  4623036865373793408  &  \\
~~~~HD ID                             & HD\,39091   & \\
~~~~TIC ID                            &  261136679 & \\
~~~~TOI ID                            &  144.01    & \\
\noalign{\vskip -3pt}
\sidehead{Spectroscopic properties}
~~~~$\teffstar$ (K)\dotfill        & $6037\pm45$   &  \citet{Ghezzi2010} \\
~~~~$\loggstar$ (cgs)\dotfill       & $4.42\pm0.03$  &  \citet{Ghezzi2010}  \\
~~~~[Fe/H] (dex)\dotfill       &  $0.08\pm0.03$ & \citet{Ghezzi2010}      \\
~~~~$v \sin i$ (\kms)\dotfill            &  $3.14\pm0.50$ & \citet{Valenti2005} \\ 
\noalign{\vskip -3pt}
\sidehead{Photometric properties}
~~~~$B$ (mag)\dotfill               & 6.25 &    \\
~~~~$V$ (mag)\dotfill               & 5.67  &     \\    
~~~~\TESS{} (mag)\dotfill            &  5.1 & TIC V7         \\
~~~~\gaia{} (mag)\dotfill            & 5.491 & \gaia{} DR2               \\
~~~~\gaia{}$_r$ (mag)\dotfill          &  
5.064 & \gaia{} DR2                 \\
~~~~\gaia{}$_b$ (mag)\dotfill          & 5.838 & \gaia{} DR2                 \\
~~~~$J$ (mag)\dotfill               & $4.87\pm0.27$ & 2MASS           \\
~~~~$H$ (mag)\dotfill               &$4.42\pm0.23$ & 2MASS           \\
~~~~$K_s$ (mag)\dotfill             & $4.241\pm0.027$  & 2MASS           \\
\noalign{\vskip -3pt}
\sidehead{Derived properties}
~~~~$\mstar$ ($\msun$)\dotfill      & $1.094\pm0.039$  & this work \\
~~~~$\rstar$ ($\rsun$)\dotfill      & $1.10\pm0.023$ &  this work         \\
~~~~$\lstar$ ($\lsun$)\dotfill      & $1.444\pm0.02$ & this work        \\
~~~~Age (Gyr)\dotfill               &  $2.98^{+1.4}_{-1.3}$  &    this work  \\
~~~~Distance (pc)\dotfill           &  $18.27\pm0.02$ & \gaia{} DR2\\
~~~~$\rhostar$ (\gcmc)\dotfill & $1.148\pm0.065$ & this work  \\
\enddata
\end{deluxetable*}

\clearpage

\begin{deluxetable*}{lcc}
\tabletypesize{\scriptsize}
\tablecaption{Parameters for the HD 39091 planetary system.\label{tab:planet}}
\startdata
\noalign{\vskip 3pt}
\hline\hline\\
{\bf Additional RV parameters} & {\bf RV offset} & \multicolumn{1}{c}{{\bf Instrument jitter}} \\
~~~ AAT (\ms) \dotfill & $32.07\pm0.86$ & \multicolumn{1}{c}{$6.7\pm0.60$} \\
~~~ HARPS pre-fix (\ms)  \dotfill & $108.51 \pm0.40$ & \multicolumn{1}{c}{$2.33\pm0.18$} \\
~~~ HARPS post-fix (\ms) \dotfill & $130.60\pm0.70$ & \multicolumn{1}{c}{$1.74\pm0.33$} \\


\noalign{\vskip 3pt}
{\bf Planet Parameters } & {\bf Planet b} &  \multicolumn{1}{c}{{\bf Planet c}} \\
~~~$P$ (days)             \dotfill    &     \AATfitPb{}    &  \multicolumn{1}{c}{\tessfitP{}}         \\
~~~$T_{p}$ (${\rm BJD}$)     \dotfill    &  \AATfittpb{}      & \multicolumn{1}{c}{-}    \\
~~~$T_{c}$ (${\rm BJD} $)     \dotfill    &  \AATfittcb{}       & \multicolumn{1}{c}{\tessfitTc{}}    \\
~~~$K$ (\ms)     \dotfill    &  \AATfittkb{}       &   \multicolumn{1}{c}{\AATfittkc{}}   \\
~~~$\sqrt{e} \cos \omega$      \dotfill    &  \AATfittsecoswb{}   &    \multicolumn{1}{c}{-}    \\
~~~$\sqrt{e} \sin \omega$      \dotfill    &  \AATfittsesinwb{}   &  \multicolumn{1}{c}{-}   \\
~~~$e$ \dotfill & \AATfiteb{} & \multicolumn{1}{c}{0} \\
~~~$\omega$ \dotfill & \AATfitwb{} &  \multicolumn{1}{c}{-} \\
~~~$T_{14}$ (hrs)  \dotfill    & -    &  \multicolumn{1}{c}{\tessfitTdur{}}    \\
~~~$\arstar$              \dotfill    &  -   & \multicolumn{1}{c}{\tessfitaor{}}  \\
~~~$\rpl/\rstar$          \dotfill    &   -  &  \multicolumn{1}{c}{\tessfitRratio{}}  \\
~~~$b \equiv a \cos i/\rstar$
                          \dotfill    &   -   &\multicolumn{1}{c}{\tessfitb{}}    \\
~~~$i_c$ (deg)              \dotfill    &  - & \multicolumn{1}{c}{\tessfitinc{}}    \\

\noalign{\vskip -3pt}
\sidehead{Derived parameters}
~~~$\mpl$      \dotfill    &   $10.02\pm 0.15$ \mjup   &  \multicolumn{1}{c}{\tessfitMp{}\mearth} \\
~~~$\rpl$ ($\rearth$)       \dotfill    & -     & \multicolumn{1}{c}{\tessfitRp{}}\\
~~~$\rhopl$ (\gcmc)       \dotfill    &  -  & \multicolumn{1}{c}{\tessfitrhop{}}   \\
~~~$\log g_p$ (cgs)       \dotfill    &  -    & \multicolumn{1}{c}{\tessfitloggp{}} \\
~~~$a$ (AU)               \dotfill    & $3.10\pm0.02$   &  \multicolumn{1}{c}{\tessfitsemi{}}  \\
~~~$T_{\rm eq}$ (K) \tablenotemark{h}        \dotfill   &  -    & \multicolumn{1}{c}{\tessfitTeq{}}   \\
~~~$\langle F_j \rangle$ ($10^{9}$ \ergscmsq)
                          \dotfill    & -  &   \multicolumn{1}{c}{\tessfitFinso{}}\\
\enddata
\end{deluxetable*}

\acknowledgments
An independent analysis of the \TESS{} data has also been reported by \citet{Gandolfi+2018}.
We acknowledge the use of TESS Alert data, 
which is currently in a beta test phase, from the TESS Science Office. 
Funding for the TESS mission is provided by NASA's Science Mission directorate.
This research has made use of the Exoplanet Follow-up Observation Program website, which is operated by the California Institute of Technology, under contract with the National Aeronautics and Space Administration under the Exoplanet Exploration Program.
CXH, JB, and MNG acknowledge support from MIT's Kavli Institute as Torres postdoctoral fellows.
AV's work was performed under contract with the California Institute of Technology / Jet Propulsion Laboratory funded by NASA through the Sagan Fellowship Program executed by the NASA Exoplanet Science Institute.
JAD and JNW acknowledge support from the Heising--Simons Foundation.
SU, FP, FB, DS, CL, DE, M.Marmier, and M.Mayor acknowledge financial support from the Swiss National Science Foundation (SNSF) in the frame work of the National Centre for Competence in Research PlanetS. DE acknowledges financial support from the European Research Council (ERC) under the European Union’s Horizon 2020 research and innovation program (project {\sc Four Aces}; grant agreement 724427).
NN acknowledges partial supported by JSPS KAKENHI Grant Number JP18H01265 and JST PRESTO Grant Number JPMJPR1775.
D.D. acknowledges support provided by NASA through Hubble Fellowship grant HSTHF2-51372.001-A awarded by the Space Telescope Science Institute, which is operated by the Association of Universities for Research in Astronomy, Inc., for NASA, under contract NAS5-26555.
We made use of the Python programming language \citep{Rossum1995} 
and the open-source Python packages
\textsc{numpy} \citep{vanderWalt2011}, 
\textsc{emcee} \citep{Foreman-Mackey2013}, 
and
\textsc{celerite} \citep{Foreman-Mackey2017}.

We acknowledge the traditional owners of the land on which the AAT stands, the Gamilaraay people, and pay our respects to elders past and present.


\textit{Facilities:} 
\facility{\TESS{}}, 
\facility{HARPS},
\facility{AAT}.




\begin{thebibliography}{}
\expandafter\ifx\csname natexlab\endcsname\relax\def\natexlab#1{#1}\fi

\bibitem[{{Bryan} {et~al.}(2018){Bryan}, {Knutson}, {Fulton}, {Lee}, {Batygin},
  {Ngo}, \& {Meshkat}}]{Bryan2018}
{Bryan}, M.~L., {Knutson}, H.~A., {Fulton}, B., {et~al.} 2018, ArXiv e-prints,
  arXiv:1806.08799

\bibitem[{{Butler} {et~al.}(2006){Butler}, {Wright}, {Marcy}, {Fischer},
  {Vogt}, {Tinney}, {Jones}, {Carter}, {Johnson}, {McCarthy}, \&
  {Penny}}]{Butler2006}
{Butler}, R.~P., {Wright}, J.~T., {Marcy}, G.~W., {et~al.} 2006, \apj, 646, 505

\bibitem[{Cameron(2012)}]{Cameron2012}
Cameron, A.~C. 2012, Nature, 492, 48

\bibitem[{{Claret}(2017)}]{Claret:2018}
{Claret}, A. 2017, \aap, 600, A30

\bibitem[{{Diego} {et~al.}(1990){Diego}, {Charalambous}, {Fish}, \&
  {Walker}}]{Diego:1990}
{Diego}, F., {Charalambous}, A., {Fish}, A.~C., \& {Walker}, D.~D. 1990, in
  \procspie, Vol. 1235, Instrumentation in Astronomy VII, ed. D.~L. {Crawford},
  562--576

\bibitem[{{Dotter} {et~al.}(2008){Dotter}, {Chaboyer}, {Jevremovi{\'c}},
  {Kostov}, {Baron}, \& {Ferguson}}]{Dotter:2008}
{Dotter}, A., {Chaboyer}, B., {Jevremovi{\'c}}, D., {et~al.} 2008, \apjs, 178,
  89

\bibitem[{{Ehrenreich} {et~al.}(2015){Ehrenreich}, {Bourrier}, {Wheatley},
  {Lecavelier des Etangs}, {H{\'e}brard}, {Udry}, {Bonfils}, {Delfosse},
  {D{\'e}sert}, {Sing}, \& {Vidal-Madjar}}]{Ehrenreich:2015}
{Ehrenreich}, D., {Bourrier}, V., {Wheatley}, P.~J., {et~al.} 2015, \nat, 522,
  459

\bibitem[{Foreman-Mackey {et~al.}(2017)Foreman-Mackey, Agol, Ambikasaran, \&
  Angus}]{Foreman-Mackey2017}
Foreman-Mackey, D., Agol, E., Ambikasaran, S., \& Angus, R. 2017, The
  Astronomical Journal, 154, 220

\bibitem[{{Foreman-Mackey} {et~al.}(2013{\natexlab{a}}){Foreman-Mackey},
  {Hogg}, {Lang}, \& {Goodman}}]{ForemanMackey:2012}
{Foreman-Mackey}, D., {Hogg}, D.~W., {Lang}, D., \& {Goodman}, J.
  2013{\natexlab{a}}, \pasp, 125, 306

\bibitem[{{Foreman-Mackey} {et~al.}(2013{\natexlab{b}}){Foreman-Mackey},
  {Hogg}, {Lang}, \& {Goodman}}]{Foreman-Mackey2013}
---. 2013{\natexlab{b}}, \pasp, 125, 306






\bibitem[{{Gaia Collaboration} {et~al.}(2018){Gaia Collaboration}, {Brown},
  {Vallenari}, {Prusti}, {de Bruijne}, {Babusiaux}, {Bailer-Jones}, {Biermann},
  {Evans}, {Eyer}, \& et~al.}]{GaiaDR2}
{Gaia Collaboration}, {Brown}, A.~G.~A., {Vallenari}, A., {et~al.} 2018, \aap,
  616, A1

\bibitem[{{Gandolfi} {et~al.}(2018){Gandolfi}, {Barragan}, {Livingston},
  {Fridlund}, {Justesen}, {Redfiel}, {Fossati}, {Mathur}, {Grziwa}, {Cabrera},
  {Garcia}, {Persson}, {Van Eylen}, {Hatzes}, {Hidalgo}, {Albrecht}, {Bugnet},
  {Cochran}, {Csizmadia}, {Deeg.}, {Eigmuller}, {Endl}, {Erikson}, {Esposito},
  {Guenther}, {Korth}, {Luque}, {Montanes Rodriguez}, {Nespral}, {Niraula},
  {Nowak}, {Patzold}, \& {Prieto-Arranz}}]{Gandolfi+2018}
{Gandolfi}, D., {Barragan}, O., {Livingston}, J., {et~al.} 2018, ArXiv
  e-prints, arXiv:1809.07573

\bibitem[{{Ghezzi} {et~al.}(2010){Ghezzi}, {Cunha}, {Smith}, {de Ara{\'u}jo},
  {Schuler}, \& {de la Reza}}]{Ghezzi2010}
{Ghezzi}, L., {Cunha}, K., {Smith}, V.~V., {et~al.} 2010, \apj, 720, 1290

\bibitem[{{G{\"u}nther} {et~al.}(2017){G{\"u}nther}, {Queloz}, {Gillen},
  {McCormac}, {Bayliss}, {Bouchy}, {Walker}, {West}, {Eigm{\"u}ller}, {Smith},
  {Armstrong}, {Burleigh}, {Casewell}, {Chaushev}, {Goad}, {Grange}, {Jackman},
  {Jenkins}, {Louden}, {Moyano}, {Pollacco}, {Poppenhaeger}, {Rauer},
  {Raynard}, {Thompson}, {Udry}, {Watson}, \& {Wheatley}}]{Guenther2017b}
{G{\"u}nther}, M.~N., {Queloz}, D., {Gillen}, E., {et~al.} 2017, MNRAS, 472,
  295

\bibitem[{G{\"u}nther {et~al.}(2018)G{\"u}nther, Queloz, Gillen, Delrez,
  Bouchy, McCormac, Smalley, Almleaky, Armstrong, Bayliss, Burdanov, Burleigh,
  Cabrera, Casewell, Cooke, Csizmadia, Ducrot, Eigmüller, Erikson, Gänsicke,
  Gibson, Gillon, Goad, Jehin, Jenkins, Louden, Moyano, Murray, Pollacco,
  Poppenhaeger, Rauer, Raynard, Smith, Sohy, Thompson, Udry, Watson, West, \&
  Wheatley}]{Guenther2018}
G{\"u}nther, M.~N., Queloz, D., Gillen, E., {et~al.} 2018, Monthly Notices of
  the Royal Astronomical Society, 478, 4720

\bibitem[{{Hadden} \& {Lithwick}(2017)}]{HaddenLithwick2017}
{Hadden}, S., \& {Lithwick}, Y. 2017, \aj, 154, 5

\bibitem[{{Jenkins} {et~al.}(2010){Jenkins}, {Caldwell}, {Chandrasekaran},
  {Twicken}, {Bryson}, {Quintana}, {Clarke}, {Li}, {Allen}, {Tenenbaum}, {Wu},
  {Klaus}, {Middour}, {Cote}, {McCauliff}, {Girouard}, {Gunter}, {Wohler},
  {Sommers}, {Hall}, {Uddin}, {Wu}, {Bhavsar}, {Van Cleve}, {Pletcher},
  {Dotson}, {Haas}, {Gilliland}, {Koch}, \& {Borucki}}]{jenkins2010}
{Jenkins}, J.~M., {Caldwell}, D.~A., {Chandrasekaran}, H., {et~al.} 2010,
  \apjl, 713, L87

\bibitem[{{Jones} {et~al.}(2002){Jones}, {Paul Butler}, {Tinney}, {Marcy},
  {Penny}, {McCarthy}, {Carter}, \& {Pourbaix}}]{Jones2002}
{Jones}, H.~R.~A., {Paul Butler}, R., {Tinney}, C.~G., {et~al.} 2002, \mnras,
  333, 871

\bibitem[{{Kempton} {et~al.}(2018){Kempton}, {Bean}, {Louie}, {Deming}, {Koll},
  {Mansfield}, {Christiansen}, {L{\'o}pez-Morales}, {Swain}, {Zellem},
  {Ballard}, {Barclay}, {Barstow}, {Batalha}, {Beatty}, {Berta-Thompson},
  {Birkby}, {Buchhave}, {Charbonneau}, {Cowan}, {Crossfield}, {de Val-Borro},
  {Doyon}, {Dragomir}, {Gaidos}, {Heng}, {Hu}, {Kane}, {Kreidberg}, {Mallonn},
  {Morley}, {Narita}, {Nascimbeni}, {Pall{\'e}}, {Quintana}, {Rauscher},
  {Seager}, {Shkolnik}, {Sing}, {Sozzetti}, {Stassun}, {Valenti}, \& {von
  Essen}}]{Kempton:2018}
{Kempton}, E.~M.-R., {Bean}, J.~L., {Louie}, D.~R., {et~al.} 2018, \pasp, 130,
  114401

\bibitem[{{Kreidberg}(2015)}]{Kreidberg(2015)}
{Kreidberg}, L. 2015, \pasp, 127, 1161

\bibitem[{{Lo Curto} {et~al.}(2015){Lo Curto}, {Pepe}, {Avila}, {Boffin},
  {Bovay}, {Chazelas}, {Coffinet}, {Fleury}, {Hughes}, {Lovis}, {Maire},
  {Manescau}, {Pasquini}, {Rihs}, {Sinclaire}, \& {Udry}}]{LoCurto2015}
{Lo Curto}, G., {Pepe}, F., {Avila}, G., {et~al.} 2015, The Messenger, 162, 9

\bibitem[{{Lopez} {et~al.}(2012){Lopez}, {Fortney}, \& {Miller}}]{Lopez:2012}
{Lopez}, E.~D., {Fortney}, J.~J., \& {Miller}, N. 2012, \apj, 761, 59

\bibitem[{{Mandel} \& {Agol}(2002)}]{MandelAgol:2002}
{Mandel}, K., \& {Agol}, E. 2002, \apjl, 580, L171

\bibitem[{{Mayor} {et~al.}(2003){Mayor}, {Pepe}, {Queloz}, {Bouchy},
  {Rupprecht}, {Lo Curto}, {Avila}, {Benz}, {Bertaux}, {Bonfils}, {Dall},
  {Dekker}, {Delabre}, {Eckert}, {Fleury}, {Gilliotte}, {Gojak}, {Guzman},
  {Kohler}, {Lizon}, {Longinotti}, {Lovis}, {Megevand}, {Pasquini}, {Reyes},
  {Sivan}, {Sosnowska}, {Soto}, {Udry}, {van Kesteren}, {Weber}, \&
  {Weilenmann}}]{Mayor2003}
{Mayor}, M., {Pepe}, F., {Queloz}, D., {et~al.} 2003, The Messenger, 114, 20



\bibitem[Owen \& Wu(2017)]{OwenWu:2017} Owen, J.~E., \& Wu, Y.\ 2017, \apj, 847, 29 



\bibitem[{{Ranalli} {et~al.}(2018){Ranalli}, {Hobbs}, \&
  {Lindegren}}]{Ranalli:2017}
{Ranalli}, P., {Hobbs}, D., \& {Lindegren}, L. 2018, \aap, 614, A30

\bibitem[{{Ricker} {et~al.}(2015){Ricker}, {Winn}, {Vanderspek}, {Latham},
  {Bakos}, {Bean}, {Berta-Thompson}, {Brown}, {Buchhave}, {Butler}, {Butler},
  {Chaplin}, {Charbonneau}, {Christensen-Dalsgaard}, {Clampin}, {Deming},
  {Doty}, {De Lee}, {Dressing}, {Dunham}, {Endl}, {Fressin}, {Ge}, {Henning},
  {Holman}, {Howard}, {Ida}, {Jenkins}, {Jernigan}, {Johnson}, {Kaltenegger},
  {Kawai}, {Kjeldsen}, {Laughlin}, {Levine}, {Lin}, {Lissauer}, {MacQueen},
  {Marcy}, {McCullough}, {Morton}, {Narita}, {Paegert}, {Palle}, {Pepe},
  {Pepper}, {Quirrenbach}, {Rinehart}, {Sasselov}, {Sato}, {Seager},
  {Sozzetti}, {Stassun}, {Sullivan}, {Szentgyorgyi}, {Torres}, {Udry}, \&
  {Villasenor}}]{ricker}
{Ricker}, G.~R., {Winn}, J.~N., {Vanderspek}, R., {et~al.} 2015, Journal of
  Astronomical Telescopes, Instruments, and Systems, 1, 014003

\bibitem[{Rossum(1995)}]{Rossum1995}
Rossum, G. 1995, Python Reference Manual, Tech. rep., Amsterdam, The
  Netherlands, The Netherlands

\bibitem[{{Seager} {et~al.}(2007){Seager}, {Kuchner}, {Hier-Majumder}, \&
  {Militzer}}]{Seager:2007}
{Seager}, S., {Kuchner}, M., {Hier-Majumder}, C.~A., \& {Militzer}, B. 2007,
  \apj, 669, 1279

\bibitem[{{Seager} \& {Mall{\'e}n-Ornelas}(2003)}]{SeagerMallenOrnelas2003}
{Seager}, S., \& {Mall{\'e}n-Ornelas}, G. 2003, \apj, 585, 1038

\bibitem[{{Spake} {et~al.}(2018){Spake}, {Sing}, {Evans}, {Oklop{\v c}i{\'c}},
  {Bourrier}, {Kreidberg}, {Rackham}, {Irwin}, {Ehrenreich}, {Wyttenbach},
  {Wakeford}, {Zhou}, {Chubb}, {Nikolov}, {Goyal}, {Henry}, {Williamson},
  {Blumenthal}, {Anderson}, {Hellier}, {Charbonneau}, {Udry}, \&
  {Madhusudhan}}]{Spake:2018}
{Spake}, J.~J., {Sing}, D.~K., {Evans}, T.~M., {et~al.} 2018, \nat, 557, 68

\bibitem[{{Triaud}(2017)}]{Triaud2017}
{Triaud}, A.~H.~M.~J. 2017, {The Rossiter-McLaughlin Effect in Exoplanet
  Research}, 2

\bibitem[Vanderburg et al.(2017)]{Vanderburg:2017} Vanderburg, A., Becker, J.~C., Buchhave, L.~A., et al.\ 2017, \aj, 154, 237 


\bibitem[{{Valenti} \& {Fischer}(2005{\natexlab{a}})}]{ValentiFischer2005}
{Valenti}, J.~A., \& {Fischer}, D.~A. 2005{\natexlab{a}}, \apjs, 159, 141

\bibitem[{{Valenti} \& {Fischer}(2005{\natexlab{b}})}]{Valenti2005}
---. 2005{\natexlab{b}}, \apjs, 159, 141

\bibitem[{van~der Walt {et~al.}(2011)van~der Walt, Colbert, \&
  Varoquaux}]{vanderWalt2011}
van~der Walt, S., Colbert, S.~C., \& Varoquaux, G. 2011, Computing in Science
  \& Engineering, 13, 22

\bibitem[{{Van Eylen} \& {Albrecht}(2015)}]{VanEylenAlbrecht2015}
{Van Eylen}, V., \& {Albrecht}, S. 2015, \apj, 808, 126

\bibitem[{{Winn}(2010)}]{Winn2010}
{Winn}, J.~N. 2010, {Exoplanet Transits and Occultations}, ed. S.~{Seager}
  (University of Arizona Press), 55--77

\bibitem[{{Winn} \& {Fabrycky}(2015)}]{WinnFabrycky2015}
{Winn}, J.~N., \& {Fabrycky}, D.~C. 2015, \araa, 53, 409

\bibitem[{{Wittenmyer} {et~al.}(2012){Wittenmyer}, {Horner}, {Tuomi}, {Salter},
  {Tinney}, {Butler}, {Jones}, {O'Toole}, {Bailey}, {Carter}, {Jenkins},
  {Zhang}, {Vogt}, \& {Rivera}}]{Wittenmyer2012}
{Wittenmyer}, R.~A., {Horner}, J., {Tuomi}, M., {et~al.} 2012, \apj, 753, 169

\bibitem[{{Zeng} {et~al.}(2016){Zeng}, {Sasselov}, \& {Jacobsen}}]{Zeng:2016}
{Zeng}, L., {Sasselov}, D.~D., \& {Jacobsen}, S.~B. 2016, \apj, 819, 127

\bibitem[{{Zhu} \& {Wu}(2018)}]{Zhu2018}
{Zhu}, W., \& {Wu}, Y. 2018, \aj, 156, 92

\bibitem[{{Zurlo} {et~al.}(2018){Zurlo}, {Mesa}, {Desidera}, {Messina},
  {Gratton}, {Moutou}, {Beuzit}, {Biller}, {Boccaletti}, {Bonavita},
  {Bonnefoy}, {Bhowmik}, {Brandner}, {Buenzli}, {Chauvin}, {Cudel}, {D'Orazi},
  {Feldt}, {Hagelberg}, {Janson}, {Lagrange}, {Langlois}, {Lannier}, {Lavie},
  {Lazzoni}, {Maire}, {Meyer}, {Mouillet}, {Peretti}, {Perrot}, {Potiron},
  {Salter}, {Schmidt}, {Sissa}, {Vigan}, {Delboulb{\'e}}, {Petit}, {Ramos},
  {Rigal}, \& {Rochat}}]{Zurlo+2018}
{Zurlo}, A., {Mesa}, D., {Desidera}, S., {et~al.} 2018, \mnras, 480, 35

\end{thebibliography}
\end{document}